\documentclass[tightenlines,a4paper,altaffilletter,twoside,prd,
               showpacs,twocolumn,floatfix,nofootinbib]{revtex4}

\usepackage[english]{babel}
\usepackage{amsmath}
\usepackage{graphicx,color}
\usepackage{epsfig,psfrag}
\usepackage{hyperref}
\usepackage{url}

\renewcommand{\url}{\nolinkurl}

\newcommand{\sthsol}{\ensuremath{\sin^2 2\theta_{12}}}     
\newcommand{\sthchooz}{\ensuremath{\sin^2 2\theta_{13}}}
\newcommand{\sthatm}{\ensuremath{\sin^2 2\theta_{23}}}
\newcommand{\deltaCP}{\ensuremath{{\delta_{\mathrm{CP}}}}}
\newcommand{\ldm}{\ensuremath{{\Delta m_{31}^2}}}          
\newcommand{\sdm}{\ensuremath{{\Delta m_{21}^2}}}

\newcommand{\emu}{\ensuremath{\nu_e \rightarrow \nu_\mu}}

\newcommand{\mue}{\ensuremath{\nu_\mu \rightarrow \nu_e}}
\newcommand{\mumu}{\ensuremath{\nu_\mu \rightarrow \nu_\mu}}
\newcommand{\mutau}{\ensuremath{\nu_\mu \rightarrow \nu_\tau}}

\newcommand{\LHC}{{\sf LHC}}
\newcommand{\ILC}{{\sf ILC}}
\newcommand{\ATLAS}{{\sf ATLAS}}
\newcommand{\CMS}{{\sf CMS}}
\newcommand{\SuperK}{{\sf Super-Kamiokande}}
\newcommand{\HyperK}{{\sf Hyper-Kamiokande}}
\newcommand{\MINOS}{{\sf MINOS}}
\newcommand{\KtoK}{{\sf K2K}}
\newcommand{\TtoK}{{\sf T2K}}
\newcommand{\NOvA}{{\sf NO$\nu$A}}
\newcommand{\CHOOZ}{{\sf CHOOZ}}
\newcommand{\DoubleChooz}{{\sf Double CHOOZ}}

\newcommand{\sgn}{\mathrm{sign}}

\begin{document}

\title{Detecting atmospheric neutrino oscillations in the ATLAS detector at CERN}
\author{Joachim Kopp}    \email[Email: ]{jkopp@mpi-hd.mpg.de}
\author{Manfred Lindner} \email[Email: ]{lindner@mpi-hd.mpg.de}
\affiliation{Max-Planck-Institut f\"ur Kernphysik \\
             Postfach 10 39 80, D-69026 Heidelberg, Germany}
\date{\today}

\begin{abstract}
  We discuss the possibility to study oscillations of atmospheric neutrinos in the
  \ATLAS\ experiment at CERN. Due to the large total detector mass, a significant
  number of events is expected, and during the shutdown phases of the \LHC,
  reconstruction of these events will be possible with very good energy and angular
  resolutions, and with charge identification. We argue that 500 live days of neutrino
  running could be achieved, and that a total of $\sim 160$ contained $\nu_\mu$ events
  and $\sim 360$ upward going muons could be collected during this time. Despite the
  low statistics, the excellent detector resolution will allow for an unambiguous
  confirmation of atmospheric neutrino oscillations and for measurements of the
  leading oscillation parameters. Though our detailed simulations show that the
  sensitivity of \ATLAS\ is worse than that of dedicated neutrino experiments, we
  demonstrate that more sophisticated detectors, e.g.\ at the \ILC, could be highly
  competitive with upcoming superbeam experiments, and might even give indications
  for the mass hierarchy and for the value of $\theta_{13}$.
\end{abstract}

\maketitle

\section{Introduction}
\label{sec:introduction}

Neutrino oscillation physics is entering the era of precision measurements, and
a plethora of new dedicated experiments are being designed, or already under
construction. The most widely discussed technologies are reactor
experiments~\cite{Ardellier:2006mn,Guo:2007ug}, superbeams~\cite{T2KProposal,
Ayres:2004js}, advanced atmospheric neutrino detectors~\cite{Nakamura:2003hk,
Rajasekaran:2004wi}, beta beams~\cite{Zucchelli:2002sa,Huber:2005jk},
and neutrino factories~\cite{Geer:1997iz,Huber:2006wb}. In this article,
we entertain the possibility to use the \ATLAS\ detector at CERN to study
oscillations of atmospheric neutrinos, an idea which has first been brought up
by F.~Vannucci~\nocite{Vannucci:PrivComm}\cite{Vannucci:PrivComm,Petcov:2005rv}.
A significant number of atmospheric neutrino interactions will take place in the 4~kt
hadronic calorimeter of \ATLAS, and can be reconstructed with excellent energy and
angular resolution during phases where the \LHC\ is not running at full luminosity.
Moreover, the magnetic field will allow for the discrimination between
neutrinos and anti-neutrinos.

To study the physics that can be done with these event, we have developped
a code that allows for the simulation of event spectra, taking into
account a full three-flavor treatment of neutrino oscillations, the finite
energy and angular resolutions of the experiment, and the detector geometry.
To derive high-level results, such as parameter sensitivities, from the simulated
data, we use a $\chi^2$ analysis including systematical uncertainties, parameter
correlations, and degeneracies.

The paper is organized as follows: After briefly reviewing the physics of
atmospheric neutrinos in Sec.~\ref{sec:atmnu}, we discuss the capabilities
of \ATLAS\ to detect atmospheric neutrinos in Sec.~\ref{sec:detector}. We
then describe the technical details of our simulations in Sec.~\ref{sec:sim},
and present the results for the sensitivity to the leading atmospheric
oscillation parameters, to the mass hierarchy, and to other three-flavor
effects, in Sec.~\ref{sec:osc}. Our conclusions will be presented in
Sec.~\ref{sec:conclusions}.

\section{Oscillations of atmospheric neutrinos}
\label{sec:atmnu}

Atmospheric neutrinos are produced by interactions of cosmic rays with the
atmosphere at a height of around 10 to 20~km above ground. The main production
reaction is
\begin{align} 
  \hspace{-0.5 cm}
  p + \textrm{Atomic Nucleus}  &\longrightarrow \pi + \textrm{Further Hadrons} \nonumber \\
                               &\hspace{1.0 cm} \begin{picture}(20,12)(0,-2)
                                                  \put(0,10){\line(0,-1){10}}
                                                  \put(0,0){\vector(1,0){20}}
                                                \end{picture} \ \ 
                                  \mu + \nu_\mu \label{eq:pi-decay} \\
                               &\hspace{2.0 cm} \begin{picture}(20,10)(0,-2)
                                                  \put(0,10){\line(0,-1){10}}
                                                  \put(0,0){\vector(1,0){20}}
                                                \end{picture} \ \
                                  e + \nu_\mu + \nu_e \nonumber
\end{align}
Due to strong $\mutau$ oscillations, the flux of upward going neutrinos is significantly
lower than that of downward going neutrinos. This leading oscillation effect has been
unambiguously detected experimentally~\cite{Fukuda:1998mi,Ashie:2004mr,Ashie:2005ik,
Aliu:2004sq,Michael:2006rx}. In contrast, subleading three-flavor effects, such as
$\mue$ oscillations driven by $\theta_{13}$ and $\ldm$, are very hard to detect
with atmospheric neutrinos because the flux ratio
$(\nu_\mu + \bar{\nu}_\mu) / (\nu_e + \bar{\nu}_e) \approx 2$
(cf.\ Eq.~\eqref{eq:pi-decay}), in combination with the close-to-maximal value of
$\theta_{23}$ leads to a cancellation of $\mue$ and $\emu$ oscillations~\cite{Peres:1999yi}.

Atmospheric neutrinos cover a wide range of energies, but their flux decreases
rapidly for $E \gtrsim 3$~GeV. The distances travelled by them before detection
range up to 12,742~km, the diameter of the Earth, for upward going neutrinos. Detailed
calculations of the atmospheric neutrino fluxes have been performed by Honda
et al.~\cite{Honda:2004yz}, Battistoni et al.~\cite{Battistoni:2002ew,
Battistoni:2003ju}, and Barr et al.~\cite{Barr:2004br}. In our study, we will
use the Honda fluxes~\cite{Honda:2004yz}.

\ATLAS\ will be sensitive mainly to the leading $\mutau$ oscillations, but due to
the difficulties associated with the reconstruction of $\nu_\tau$ interactions,
the most important oscillation channel will be $\mumu$ disappearance. Neglecting
subleading effects driven by $\sdm$, the corresponding oscillation probability
in matter is given by~\cite{Akhmedov:2004ny}:
\begin{multline}
  P_{\mu\mu} = 1 - \sin^4\theta_{23} \sin^2 2\theta_{13}^m \sin^2 C_{13}\Delta \\
       - \frac{1}{2} \sthatm \big[ 1 - \cos(1+A)\Delta \cos C_{13}\Delta \\
             + \cos 2\theta_{13}^m \sin(1+A)\Delta \sin C_{13}\Delta \big]
  \label{eq:Pmumu}
\end{multline}
Here, we have used the notation from~\cite{Akhmedov:2004ny}:
$\Delta = \ldm L/4E$ is the oscillation phase,
$A = 2 E V/\ldm$ describes matter effects,
$\sin 2\theta_{13}^m = C_{13}^{-1} \sin 2\theta_{13}$ is
the mixing angle in matter, and $C_{13}$ is given by
$C_{13} = [\sthchooz + (\cos 2\theta_{13} - A)^2]^{1/2}$.
The first line in Eq.~\eqref{eq:Pmumu} describes the disappearance due to
matter-enhanced $\mue$ oscillations, while the other terms correspond to
$\mutau$ oscillations. A generic three-flavor effect is the strong influence
of matter on the $\mumu$ and $\mutau$ channels in some regions of the parameter
space~\cite{Gandhi:2004md,Gandhi:2004bj}. In the vacuum case, i.e.\ for
$A=0$, $C_{13}=1$, and $\theta_{13}^m = \theta_{13}$, Eq,~\eqref{eq:Pmumu}
reduces to
\begin{multline}
  P_{\mu\mu} = 1 - \big[ \sin^4\theta_{23} \sin^2 2\theta_{13}
  - \cos^2\theta_{13} \sin^2 2\theta_{23} \big] \sin^2\Delta,
\end{multline}
i.e.\ to two-flavor oscillations with a mixing angle close to $\theta_{23}$.
For the neutrino oscillation parameters, we adopt the following best-fit values
in our simulations (see e.g.\ ~\cite{Maltoni:2004ei,Fogli:2005cq,Bahcall:2004ut,
Bandyopadhyay:2004da}):
\begin{gather}
  \begin{split}
    \sthsol         &= 0.79, \\
    \sthchooz       &= 0.12, \\
    \sthatm         &= 1.0, \\
    \delta_{\rm CP} &= 0.0, \\
    \sdm            &= 8.1 \cdot 10^{-5} \ \textrm{eV}^2, \\
    \ldm            &= 2.2 \cdot 10^{-3} \ \textrm{eV}^2.
  \end{split}
  \label{eq:true-values}
\end{gather}

\section{Reconstruction of neutrino interactions in ATLAS}
\label{sec:detector}

The \ATLAS\ detector at CERN has an onion shell structure, similar to that of most
other modern collider experiments: The inner high-resolution tracking
detectors are surrounded by the electromagnetic and hadronic calorimeters, large
superconducting magnets, and a muon tracking system~\cite{Armstrong:1994it}. \ATLAS\ has
a total mass of 7~kt, but part of it is attributed to the non-active support structure
of the experiment. Reconstruction of neutrino interactions will only
be possible if the energy and direction of the secondary charged lepton
can be seen, and for a good energy and angular resolution, it is desirable
to also reconstruct the energies of the other interaction products.
For $\nu_\mu$ interactions, this requires the interaction to take place
inside the 4~kt hadronic calorimeter, so that the muons will travel
through the whole muon system. Furthermore, the neutrino
must have sufficient energy to yield a sizeable signal. We estimate that
a threshold of 1.5~GeV should be realistic. For $\nu_e$ and $\nu_\tau$, the
track of the secondary charged lepton can only be seen with good resolution
if the interaction takes place in the inner tracking detectors, which, however,
have negligible mass. Energy reconstruction for $\nu_e$ and $\nu_\tau$
should in principle be possible also for interactions taking place further outside,
but the resolution would be quite poor since only information from the hadronic
calorimeter could be used. Therefore, we do not take $\nu_e$ and $\nu_\tau$
into account at all. We do, however, include so-called upward going muons,
i.e.\ muons created by neutrino interactions in the rock below the detector.
More precisely, we consider muons coming from the zenith angle range
$-1 \leq \cos\theta \leq -0.1$ to be induced by neutrinos. Note that \ATLAS,
being a magnetized detector, is able to determine the muon charge and
can thus discriminate between neutrinos and anti-neutrinos.

A very important issue for the detection of atmospheric neutrinos in \ATLAS\
is triggering. The characteristic signature of an atmospheric neutrino is
a high-energy charged lepton track originating from a vertex inside the detector,
without any visible ingoing particle. During normal \LHC\ operation at full
luminosity, such signals will be cloaked by the pile-up of $pp$ interaction
products, and even if they could be seen, they might still be confused with
decay products of neutral hadrons. Besides, the \ATLAS\ trigger will only
be sensitive during the bunch crossings, because the time between them is
required to read out the accumulated data. For these reasons, \ATLAS\ can
be used for neutrino physics only during the $\sim 200$~days per year where
the \LHC\ is not running in collider mode~\cite{LHCSchedule}. This number
arises from the 14~week winter shutdown, and from regular short maintenance
shutdowns throughout the year. Moreover, even during \LHC\ operation, the $pp$
interaction rate in the detector will be sufficiently low during the ramp-up and
ramp-down phases. Of course, also the \ATLAS\ detector will require maintenance,
so that only part of the aforementioned time windows will be available for neutrino
physics. If $\sim 100$~days of neutrino running per year are feasible, this would
yield $\sim 160$ contained events and $\sim 360$ upward going muons within
5~years. We will show in Sec.~\ref{sec:osc} that this is sufficient to
detect neutrino oscillations and to perform measurements
of the oscillation parameters.

Note that the number of neutrino interactions expected in \CMS\ should be even
larger than that in \ATLAS, since \CMS\ has a total mass of 12~kt.~\cite{unknown:1994pu}.
However, most of this mass is concentrated in the massive iron return yokes of
the magnet, so that all non-muonic interaction products will be scattered
several times before reaching the active detector components. This will
inhibit a reliable reconstruction of the primary neutrino properties, and
hence we will not consider \CMS\ further in this article.

Since the leading oscillation channel for atmospheric neutrinos is $\mutau$, and
the $\tau$ has a 17\% chance of decaying into $\mu \nu_\mu \nu_\tau$~\cite{Yao:2006px},
one has to ask the question whether $\nu_\tau$ interactions, followed by a leptonic
$\tau$ decay, can be clearly separated from simple charged current $\nu_\mu$
interactions. However, this should be possible, since typically, the muon from the
$\tau$ decay will have a relatively low energy, which is most probably below
the threshold used in the analysis.

\section{Simulation of atmospheric neutrinos}
\label{sec:sim}

To study the prospects of doing neutrino oscillation physics with \ATLAS,
we have developed a simulation code for the calculation of event rates
as a function of the neutrino energy and zenith angle, and for the $\chi^2$
analysis of the simulated data. We obtain the binned event rates
by folding the initial neutrinos fluxes $\Phi_{f^\prime}\big( E_\nu^k, L(\theta_\nu^l) \big)$,
\cite{Honda:2004yz}, the three-flavor oscillation probabilities in matter
$P\big( f^\prime \rightarrow f, E_\nu^k, L(\theta_\nu^l), \vec{\Theta} \big)$,
the cross sections $\sigma_f(E_\nu^k)$~\cite{Messier:1999kj,Paschos:2001np},
and a detector response function $\tilde{R}^{ij}(E_\nu^k, \theta_\nu^l)$
according to the formula
\begin{multline*}
  N_f^{ij} = \mathcal{N} \, \sum_{k,l} \tilde{R}^{ij}(E_\nu^k, \theta_\nu^l) \cdot \sigma_f(E_\nu^k) \\
   \cdot \sum_{f^\prime = e,\mu,\tau} \!\!\! P\big( f^\prime \rightarrow f, E_\nu^k, L(\theta_\nu^l),
                                           \vec{\Theta} \big)
   \, \Phi_{f^\prime}\big( E_\nu^k, L(\theta_\nu^l) \big).
\end{multline*}
In this expression, $E_\nu^k$ and $\theta_\nu^l$ denote the neutrino energy
and zenith angle at the sampling point with indices $(k, l)$, while $(i,j)$ stands
for the binning used in the analysis. The number of bins is roughly chosen according
to the energy and angular resolutions (see Tab.~\ref{tab:scenarios}).%
\footnote{This binning could not be employed in the analysis of real data, since
the event numbers in each bin would be very small, but for our purposes, it gives
a good estimate of the information that is contained in the data.}
$L(\theta_\nu^l)$ is the distance travelled by neutrinos coming from the direction
$\theta_\nu^l$, and $\vec{\Theta}$ represents the vector of oscillation paramaters
(including the matter potential), i.e.\ $\vec{\Theta} = (\theta_{12},
\theta_{13}, \theta_{23}, \deltaCP, \sdm, \ldm, V)$. The cross
sections used in our code cover an energy range from 100~MeV to 1~TeV.
For neutrinos with higher energies (up to 10~TeV), which can contribute
only to the upward going muon sample, we extrapolate the cross sections
by making the assumption that $\sigma(E)/E$ is constant at such high
energies. We only use total charged current cross sections.
The detector response function $\tilde{R}^{ij}(E_\nu^k, \theta_\nu^l)$
determines which fraction of neutrinos with initial energy $E_\nu^k$
and zenith angle $\theta_\nu^l$ is reconstructed into bin $(i,j)$.
Since the detector response to neutrinos has not been studied in full
detector Monte Carlo simulations yet, we resort to approximating $\tilde{R}$
by a double Gaussian resolution function, multiplied with an efficiency
factor $\epsilon(E_\nu, \theta_\nu)$:
\begin{multline*}
  \tilde{R}^{ij}(E_\nu^k, \theta_\nu^l) =
     \epsilon(E_\nu, \theta_\nu^l) 
     \cdot \frac{1}{Z_E} \exp\left(- \frac{(E_r^i - E_\nu)^2}{2\sigma_E^2 (E_\nu)} \right)  \\
     \cdot \frac{1}{Z_\alpha} \int_0^{2\pi} d\phi_r
               \exp\left(- \frac{\alpha^2(\theta_r^j, \phi_r, \theta_\nu^l)}
                                            {2\sigma_\alpha^2 (E_\nu)} \right).
\end{multline*}
Here, $\sigma_E (E_\nu)$ and $\sigma_\alpha (E_\nu)$ denote the
energy and angular resolutions, respectively.
$\alpha(\theta_r^j, \phi_r, \theta_\nu^l)$ is the angle between
the initial neutrino direction, parameterized by the zenith angle
$\theta_\nu^l$, and the reconstructed direction, which is defined
by the zenith and azimuthal angles $\theta_r^j$ and $\phi_r$;
$Z_E$ and $Z_\alpha$ are normalization factors which ensure that
the total number of events is conserved in the ``smearing'' of
the spectrum. Due to lack of Monte Carlo results on the reconstruction
efficiency in \ATLAS, we omit $\epsilon(E_\nu, \theta_\nu)$ for
contained events. For upward going muons, we use $\epsilon$ to scale
the flux according to the target volume, which varies with energy and
zenith angle (see appendix~\ref{sec:geometry} for details).

For the resolutions, we use estimates based on the experience of
the \SuperK\ collaboration~\cite{Ishitsuka:2004gd,Ashie:2005ik} and on the
information from the \ATLAS\ proposal~\cite{Armstrong:1994it}.
In \SuperK, where only the secondary muon is seen, the energy resolution
for contained Multi-GeV events is $17\%$, and the angular resolution is
$17^\circ$ above 1.5~GeV. The \ATLAS\ calorimeter measures the full deposited
energy with a resolution of order $\sigma_E/E \sim 50\% / \sqrt{E / {\rm GeV}}$,
while the muon system has an energy resolution on the per cent level. The
angular resolution of the muon system is excellent because track resconstruction
is possible, while the directional information from the hadronic calorimeter
is severely limited by its rough segmentation. We will assume that the excellent muon
reconstruction, in conjuction with the information from the hadronic calorimeter,
will yield an overall neutrino energy resolution of 5\%, and an angular
resolution of $7^\circ$ for contained events. Let us emphasize that these
numbers are only estimates, and for more reliable numbers, detailed detector
Monte Carlo simulations are indispensable. For upward going muons, no energy
reconstruction is possible, and the angular resolution is essentially given
by the average difference between the neutrino and muon directions. Based
on~\cite{Ashie:2005ik}, we take $\sigma_\alpha = 5^\circ$ for upward going
muon events.

To account for the case that our estimates are not conservative enough, we will
also consider a scenario where the resolutions are as poor as in \SuperK,
and, in addition, the live time is taken to be only 250~days. On the other
hand, we will also discuss a very optimistic setup with $\sigma_E/E = 2\%$
and $\sigma_\alpha = 2^\circ$ for contained events. In this setup, we also
lower the energy threshold to 0.3~GeV and increase the exposure time to
2,000~days. These assumptions will most probably not apply to \ATLAS, since
they would correspond to a situation where neutrino reconstruction is
possible even during normal \LHC\ operation, and the reconstruction efficiency
is almost 100\%. However, they may be interesting in the context of future
projects such as the proposed \ILC\ detectors~\cite{ILC-WWStudy}. In an \ILC\
experiment, no pile-up of hadronic events will occur, so neutrino detection
is not restricted to the maintenace phases of the accelerator. The parameters
of our three scenarios are summarized in Tab.~\ref{tab:scenarios}

\begin{table}
  \centering
  \begin{ruledtabular}
    \begin{tabular}{lrrr}
        & \bf{\hfill\ATLAS\hfill}     & \bf{\hfill\ATLAS\hfill} & \bf{\hfill\ILC\hfill} \\
        & \bf{\hfill Realistic\hfill} & \bf{\hfill Cons.\hfill} &                       \\ \hline
      Running time                         &   500~d      & 250~d      & 2,000~d   \\
      Energy threshold                     &   1.5~GeV    & 1.5~GeV    & 0.3~GeV   \\
      $\sigma_E/E$ (contained events)      &     5\%      & 17\%       & 2\%       \\
      $\sigma_\alpha$ (contained events)   &    $7^\circ$ & $17^\circ$ & $2^\circ$ \\
      $\sigma_\alpha$ (upward muons)       &    $5^\circ$ & $5^\circ$  & $5^\circ$ \\
      $E$ bins                             &    30        & 20         & 90        \\
      $\cos\theta$ bins (contained events) &    36        & 10         & 80        \\
      $\cos\theta$ bins (upward muons)     &    27        & 27         & 27        \\
    \end{tabular}
  \end{ruledtabular}
  \caption{Properties of the realistic and conservative \ATLAS\ scenarios, andof
           an assumed \ILC\ detector.}
  \label{tab:scenarios}
\end{table}

To analyze the simulated event spectra, we use a $\chi^2$ fit which distinguishes
neutrinos from anti-neutrinos, i.e.\ we assume 100\% charge identification efficiency.
For each species, we define a $\chi^2$ function of the form
\begin{multline}
  \chi^2_{\rm stat} = \sum_{i,j} 2 \big[T_{ij}(\vec{\Theta}_{\rm fit}, \vec{a})
                                      \ -\ N_{ij}(\vec{\Theta}_{\rm true}) \big] \\
     + 2 \, N_{ij}(\vec{\Theta}_{\rm true})
                   \ln \bigg( \frac{N_{ij}(\vec{\Theta}_{\rm true})}
                                    {T_{ij}(\vec{\Theta}_{\rm fit}, \vec{a})} \bigg) \\
     + \chi^2_{\rm pull,osc} + \chi^2_{\rm pull,sys}.
  \label{eq:chi2}
\end{multline}
Here, $N_{ij}(\vec{\Theta}_{\rm true})$ is the ``observed'' event rate in bin $(i,j)$ for
the ``true'' oscillation parameters
$\vec{\Theta}_{\rm true} = \left(\theta_{12}, \theta_{13}, \theta_{23}, \deltaCP,
                                 \sdm, \ldm, V \right)$,
and $T_{ij}(\vec{\Theta}_{\rm fit}, \vec{a})$ is the event rate that would be expected
for the hypothesized parameters $\vec{\Theta}_{\rm fit}$, and for biases $\vec{a}$ arising
from systematical errors in the experiment. We have introduced pull terms
\begin{equation}
  \chi^2_{\rm pull,osc} = \sum_k \frac{(\Theta_{k, {\rm fit}} - \Theta_{k, {\rm true}})^2}
                                                                        {\sigma_{\Theta_k}^2},
\end{equation}
to account for external input on the oscillation parameters. $\sigma_{\Theta_k}$
determines how strongly a fit value far from the externally given one is disfavoured.
Pull terms are provided only for those parameters which are marginalized over in the
fit. We take relative $1\sigma$ uncertainties of 10\% for $\theta_{12}$ and $\theta_{23}$, 5\%
for $\sdm$, and 30\% for $\ldm$. $\theta_{13}$ is assigned an absolute uncertainty
of $10^\circ$.

Pull terms for systematical biases are similar to those for the oscillation parameters:
\begin{equation}
  \chi^2_{\rm pull,sys} = \sum_k \frac{a_k^2}{\sigma_{a_k}^2}.
\end{equation}
They disfavor fit values $a_k$ which are further from zero than can
be expected from the systematical uncertainties $\sigma_{a_k}$. We have defined $\vec{a}$
such that the case of vanishing systematical errors corresponds to $\vec{a} = 0$.
The various types of systematical errors we consider, are summarized
in Tab.~\ref{tab:sys}. Besides the usual normalization errors we allow also for
``tilts'' in the event spectrum, which are a simple way of accounting for energy or
angle dependent biases.

Note that many of the technical details of our simulations and of our statistical
analysis procedure follow ideas that have been previously realized in the GLoBES
code~\cite{Huber:2004ka,Huber:2007ji}.

\begin{table}
  \begin{ruledtabular}
    \begin{tabular}{lr}
      \bf Error Type                                         & $\mathbf{\sigma_a}$ \\ \hline
      Overall normalization for contained events             &  20\%  \\
      Relative normalization for anti-neutrinos              &   5\%  \\
      Normalization for upward going muon events             &  20\%  \\
      Tilt of $E$ spectrum                                   &   5\%  \\
      Tilt of $\theta$ spectrum for contained events         &  10\%  \\
      Tilt of $\theta$ spectrum for upward going muon events &   2\%  \\
    \end{tabular}
  \end{ruledtabular}
  \caption{Systematical errors in \ATLAS.}
  \label{tab:sys}
\end{table}

\section{Simulation results}
\label{sec:osc}

\begin{figure}[tbp]
  \begin{center}
    \includegraphics[width=8 cm]{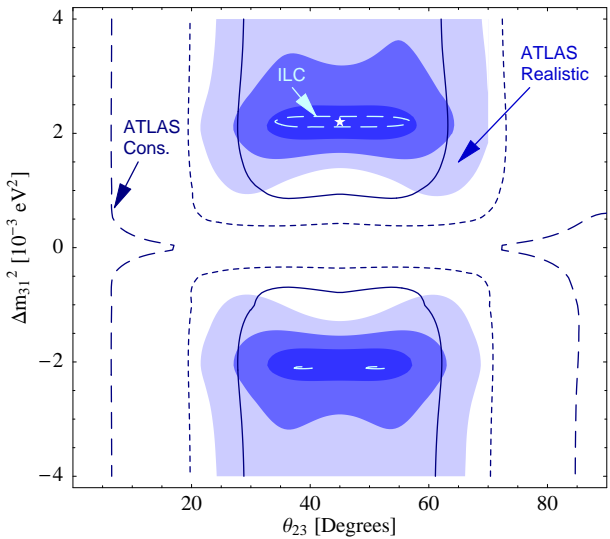}
  \end{center}
  \vspace{-0.5 cm}
  \caption{Sensitivity of \ATLAS\ and of a possible \ILC\ detector to the leading
    atmospheric oscillation parameters $\theta_{23}$ and $\ldm$. The shaded regions
    correspond to the $1\sigma$, $2\sigma$, and $3\sigma$ confidence levels for the
    realistic scenario, while the dark contours correspond to the conservative case.
    For the \ILC\ scenario (light-colored contours), we show only the $3\sigma$
    contour for clarity. A normal mass hierarchy has been assumed in the plot, but
    we have checked that all results except the sensitivity to the mass hierarchy,
    are similar for the inverted hierarchy (see text for details).}
  \label{fig:th23m31-atlas}
\end{figure}

\begin{figure}[tbp]
  \begin{center}
    \includegraphics[width=8 cm]{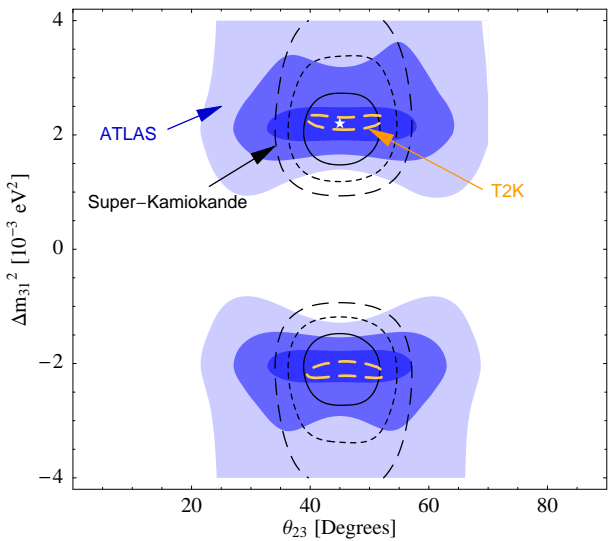}
  \end{center}
  \vspace{-0.5 cm}
  \caption{Sensitivity of \ATLAS\ to the leading atmospheric oscillation parameters
    $\theta_{23}$ and $\ldm$ in comparison with a global fit to existing atmospheric
    neutrino data (dominated by \SuperK)~\cite{Maltoni:2004ei,Schwetz:2006dh},
    and with the expected performance of \TtoK. For the latter experiment, we show
    only the $3\sigma$ contour for clarity. A normal mass hierarchy was assumed in
    the plot, but we have checked that the results for the inverted hierarchy are
    analogous.}
  \label{fig:th23m31-allexps}
\end{figure}

In this section, we present our main results on the sensitivity
of \ATLAS\ to the leading atmospheric oscillation parameters and to three-flavor
effects. In Fig.~\ref{fig:th23m31-atlas}, we compare the expected confidence
regions in the $\theta_{23}$--$\ldm$ plane for the three scenarios introduced
in Sec.~\ref{sec:sim}, and in Fig.~\ref{fig:th23m31-allexps} we relate them
to the results of a global three-flavor fit to existing atmospheric neutrino
data~\cite{Maltoni:2004ei,Schwetz:2006dh}, and to the expected performance of
\TtoK~\cite{Itow:2001ee,Huber:2002mx,Ishitsuka:2005qi}. From the plots, we
first notice that in all three scenarios, neutrino oscillations can be
confirmed at better than $3\sigma$, i.e.\ the case $\theta_{23} = 0$, $\ldm = 0$
can be ruled out. However, in the conservative case, the sensitivity is only
marginal due to the limited statistics and resolutions, and could be easily
spoiled by unfavorable statistical fluctuations in the real experiments.

The precision with which the mixing angle can be determined is poor in all three
scenarios: Even in the optimistic \ILC\ case, the uncertainty at
$3\sigma$ is about $\pm 30\%$, which is still slightly worse than the current bound from
\SuperK, \KtoK, and \MINOS ($\sim 23\%$, cf.~\cite{Maltoni:2004ei}). A measurement
of $\theta_{23}$ in \ATLAS\ is not competitive at all. The sensitivity to
$\theta_{23}$ in atmospheric neutrinos comes mainly from the up-down
asymmetry of the neutrino flux, which can only be measured accurately with good
statistics. The excellent resolution of \ATLAS\ cannot compensate its small mass here.

The situation is better for $\ldm$: For the realistic scenario, \ATLAS\ shows a
performance similar to \SuperK, at least at low confidence levels. The
$3\sigma$ contour, however, extends much further to large $\ldm$. In the
\ILC\ scenario, the sensitivity to $\ldm$ is even comparable to that of \TtoK.
An atmospheric neutrino oscillation experiment is sensitive to the mass
squared difference mainly through the shape of the zenith angle spectrum
at around $\cos\theta \approx 0$. These directions correspond to baselines
below the first oscillation maximum, where the depletion of the flux is just
setting in. Wash-out due to poor detector resolutions can greatly limit
this measurement, so \ATLAS-like detectors have a considerable advantage here.

Turning to the discovery potential for generic three-flavor effects, we remark,
that although a very large value of $\sthchooz = 0.12$ was assumed in
Fig.~\ref{fig:th23m31-atlas} and Fig.~\ref{fig:th23m31-allexps}, the wrong
hierarchy solution~\cite{Minakata:2001qm,Barger:2001yr} cannot be fully ruled out
at $3\sigma$ even by an advanced \ILC\ detector. In the two \ATLAS\ scenarios,
there is no sensitivity to the hierarchy at all. If the true hierarchy is
inverted, matter effects are shifted to the anti-neutrino channels, where they
are cloaked due to the smaller anti-neutrino cross sections, and we have checked
that in this case, even the \ILC\ scenario has only a $1\sigma$ discovery
potential. One might have hoped that matter effects in the $\mumu$ channel
could have helped to resolve this degeneracy~\cite{Gandhi:2004md,Gandhi:2004bj},
but it turns out that the event numbers are too small in the relevant
energy range of several GeV.

\begin{figure}[tbp]
  \begin{center}
    \includegraphics[width=9 cm]{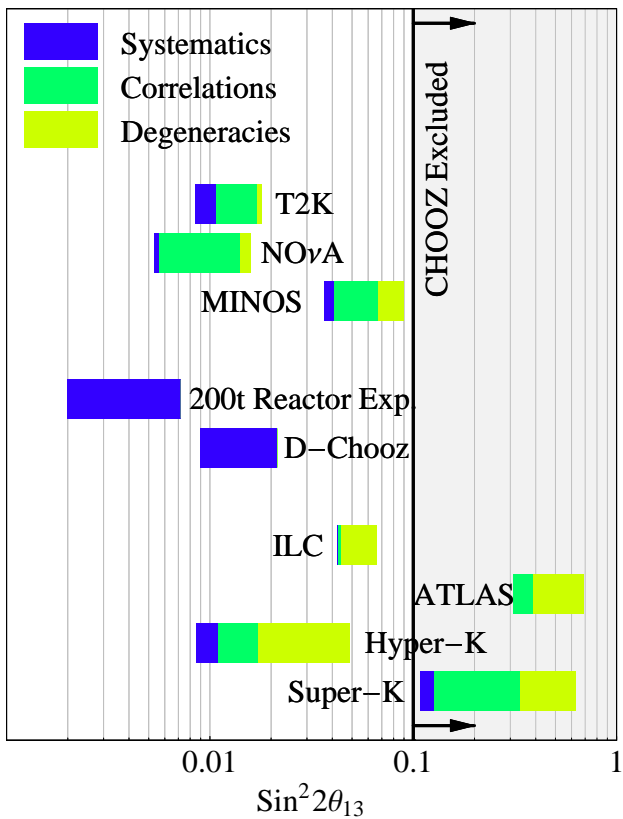}
  \end{center}
  \vspace{-0.5 cm}
  \caption{Sensitivity of various neutrino oscillation experiments to $\sthchooz$.
  The colored bars reflect the limitations of the sensitivity due to systematical
  errors, parameter correlations, and degeneracies. The right edges of the bars
  correspond to the sensitivity which is expected in reality.}
  \label{fig:th13bars}
\end{figure}

For the same reason, we also expect a poor performance in the investigation
of other three-flavor effects. To demonstrate this, we have plotted in
Fig.~\ref{fig:th13bars} the sensitivity of the realistic \ATLAS\ scenraio
and several other experiments to $\sthchooz$. As examples for reactor experiments,
we show \DoubleChooz~\cite{Ardellier:2006mn}, and a more advanced setup
with a 200~t far detector~\cite{Huber:2006vr}. On the accelerator side,
we show simulation results for \MINOS~\cite{Ables:1995wq}, \TtoK~\cite{T2KProposal},
and \NOvA~\cite{Ayres:2004js,Yang:2004}. The sensitivity to $\sthchooz$ is
defined as the limit, which a specific experiment can set to $\sthchooz$,
assuming the true value is zero. The left edges of the bars correspond to 
the hypothetical case that only statistical errors are present in the
experiment. The blue, green, and yellow bars reflect the limitations of
the sensitivity due to systematical uncertainties, parameter correlations,
and the $\sgn(\ldm)$ degeneracy.%
\footnote{The octant degeneracy~\cite{Barger:2001yr} is irrelevant here
since we have assumed $\theta_{23}^{\rm true} = \pi/4$, and the intrinsic
degeneracy~\cite{Burguet-Castell:2001ez} appears only in high statistics
experiments such as a neutrino factory.}
We can read off from Fig.~\ref{fig:th13bars} that atmospheric neutrino
experiments can in general not compete with reactor and accelerator setups.
Only \HyperK\ and the \ILC\ scenario could improve the
\CHOOZ\ bound, though not as much as dedicated experiments.
It is interesting to note that the \ILC\ scenario has a
significantly better sensitivity than \SuperK, although it has only
$\sim 1,350$ contained $\nu_\mu$ events and $\sim 1,720$ upward going
muons, while \SuperK\ has detected $\sim 4,500$ fully contained $\mu$-like
events and $\sim 2,250$~upward going muons, plus a large number
of $e$-like events and partially contained events~\cite{Ashie:2005ik}.
This comparison confirms the statement that, to a certain degree, 
excellent resolutions can compensate for low statistics in atmospheric
neutrino experiments~\cite{Petcov:2005rv}.

\section{Conclusions}
\label{sec:conclusions}

We have shown that atmospheric neutrino interactions are an interesting by-product
of operating the \ATLAS\ detector. Although reconstruction of these events
is restricted to phases where the \LHC\ is not running, or only running at low
luminosity, the number of events, in conjunction with the expected good
energy and angular resolutions, is sufficient to confirm atmospheric neutrino
oscillations and to measure the leading oscillation parameters $\theta_{23}$
and $\ldm$. However, the precision of these measurements will probably not
be competitive to that of existing and upcoming dedicated experiments.
Under more optimistic assumptions, which might be realized at future \ILC\
detectors, the errors on $\theta_{23}$ and $\ldm$ can become comparable
to those of \TtoK, and, if $\theta_{13}$ is very large, the $\sgn(\ldm)$
degeneracy might be resolved at the $2\sigma$ level.

To study the reconstruction capabilities and sensitivities of \ATLAS\
in more detail, it will be necessary to perform detailed detector
Monte Carlo simulations. We believe this to be definitely worthwhile,
since a measurement of the atmospheric neutrino oscillation parameters
with the technology of \ATLAS\ would be a very interesting result in itself,
even if the final error bars should be larger than those of other experiments.
Moreover, the exploration of neutrino interactions in \ATLAS\ would
provide valuable experience that could be of great interest in the design of
future collider experiments and neutrino detectors.

\section*{Acknowledgments}

We would like to thank E.~Kh.~Akhmedov, S.~Goswami, K.~A.~Hochmuth, P.~Huber,
J.~Schmaler, W.~Smith, and especially F.~Vannucci for useful discussions. We are
especially grateful to T.~Schwetz-Mangold for kindly providing the results of the
global fit to atmospheric neutrino data from~\cite{Maltoni:2004ei,Schwetz:2006dh}
in machine-readable form. JK would like to acknowledge support from the
Studienstiftung des Deutschen Volkes.

\appendix
\section{Geometry of upward going muons in \ATLAS}
\label{sec:geometry}

\begin{figure}[tbp]
  \vspace{0.3 cm}
  \begin{center}
    \psfrag{Beam}{\color{black} Muons}
    \psfrag{Lmu}{\color{blue} $L_\mu$}
    \psfrag{theta}{\color{blue} $\theta$}
    \psfrag{phi}{\color{blue} $\phi$}
    \psfrag{eta}{\color{blue} $\eta$}
    \psfrag{l}{\color{blue} $\ell$}
    \psfrag{2r}{\color{blue} 2r}
    \psfrag{xx}{\color{blue} x}
    \psfrag{yy}{\color{blue} y}
    \psfrag{zz}{\color{blue} z}
    \hspace{2 cm}
    \includegraphics[width=7 cm]{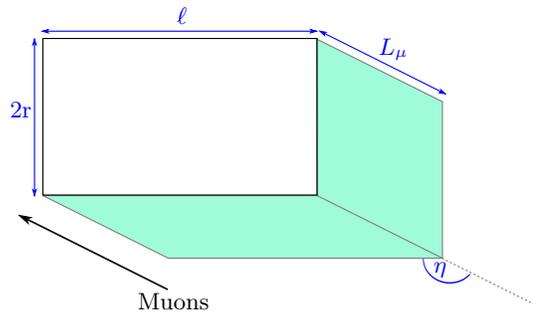}
  \end{center}
  \vspace{-0.2 cm}
  \caption{Geometry of upward going muons. The white rectangle schematically
           shows the \ATLAS\ experiment as seen from a direction orthogonal
           to the detetor axis and to the muon direction. The shaded region
           represents the effective target volume $V$.}
  \label{fig:atlas-geom}
\end{figure}

As mentioned in Sec.~\ref{sec:sim}, the target volume for upward going
muon events depends on the energy and on the zenith and azimuthal angles.
It can be calculated from geometrical arguments as follows. We first
integrate the inverse of the Bethe-Bloch formula~\cite{Eidelman:2004wy,
Groom:2001kq} to obtain the ``muon range'' $L_\mu$, i.e.\ the average
distance that a muon with energy $E_\mu$ can travel in rock. We then use
Fig.~\ref{fig:atlas-geom} to calculate the effective target volume
$V$ in which a neutrino must interact to induce an upward going muon
event in the detector. The white rectangle in the diagram represents
the cylindrical \ATLAS\ detector as seen from a direction orthogonal to
the detector axis and to the muon direction, so that the angle $\eta$ between
these two vectors lies in the drawing plane. The effective target volume,
shown by the shaded region in Fig.~\ref{fig:atlas-geom}, is given by
\begin{equation}
  V =  L_\mu |\sin\eta| \cdot \ell \cdot 2r + L_\mu |\cos\eta| \cdot \pi r^2.
\end{equation}
We can calculate $\eta$ from the azimuthal angle $\phi$ and the zenith angle
$\theta$ by
\begin{equation}
  \cos\eta = \sin\theta \cos\phi.
\end{equation}
Averaging over $\phi$ yields
\begin{align}
  \bar{V} &= \frac{L_\mu}{2\pi} \int_0^{2\pi} \!\!\!\! d\phi
       ( 2r\ell \sqrt{1 - \sin^2\theta \cos^2\phi} + \pi r^2 |\sin\theta \cos\phi| ) \nonumber\\
          &= \frac{4 r \ell L_\mu}{\pi} E(\sin\theta) + 2 r^2 L_\mu |\sin\theta|,
\end{align}
with $E(\cdot)$ being the complete elliptic integral of the second
kind~\cite{Abramowitz:1964}.

When evaluating this expression, we assume a horizontally zylindrical
geometry for \ATLAS, with a length of 42~m and a radius of 11~m. For the
geometrical arguments, the muon direction is taken to be identical
to that of the primary neutrino, which is justified by the fact that upward
going muons typically have very high energies $\gg 1$~GeV. The muon
energy is calculated from the neutrino energy according to the empirical
formula
\begin{align}
  E_\mu = E_\nu \left( e^{ 5.255 - 1.819 \log_{10} (E_\nu / {\rm MeV}) }
              + 0.3298 \right),
\end{align}
which we have obtained by fitting $E_\mu/E_\nu$ in a sample of 1,000 upward
going muon events in \SuperK, simulated with the NUANCE event
generator~\cite{Casper:2002sd}.

\bibliography{atlas}

\end{document}